\documentclass[10pt,journal]{IEEEtran}

\usepackage[mathscr]{eucal}
\usepackage[tight,footnotesize]{subfigure}
\usepackage[cmex10]{amsmath}
\usepackage{cite}
\usepackage{afterpage,layout,longtable,varioref,verbatim}
\usepackage{algorithmic}
\usepackage{array}
\usepackage{graphics}
\usepackage{color}
\usepackage{latexsym}
\usepackage{epsf,psfrag}
\usepackage{epsfig}
\usepackage{amssymb}
\usepackage{textcomp}
\usepackage{amsmath}
\usepackage{bm}

\newtheorem{theorem}{Theorem}[section]

\newtheorem{lemma}{Lemma}[section]
\newtheorem{corollary}{Corollary}[theorem]

\newtheorem{definition}{Definition}[section]

\newcommand{\eg}{\textit{e.g.}}
\newcommand{\ie}{\textit{i.e.}}

\definecolor{dkr}{rgb}{0.6,0.2,0.2}

\definecolor{light-gray}{gray}{0.85}
\definecolor{BrickRed}{RGB}{150,25,14}

\newcommand{\Amsc}{\mathscr{A}}

\newcommand{\Cmsc}{\mathscr{C}}
\newcommand{\Emsc}{\mathscr{E}}

\newcommand{\Gmsc}{\mathscr{G}}

\newcommand{\Nmsc}{\mathscr{N}}

\newcommand{\Rmsc}{\mathscr{R}}
\newcommand{\Smsc}{\mathscr{S}}
\newcommand{\SmscA}{\mathscr{S}_{\mbox{\scriptsize A}}}

\newcommand{\SmscF}{\mathscr{S}_{\mbox{\scriptsize F}}}

\newcommand{\Smsco}{\mathscr{S}_{\mbox{\scriptsize o}}}

\newcommand{\Vmsc}{\mathscr{V}}

\newcommand{\Xmsc}{\mathscr{X}}
\newcommand{\Xmsco}{\mathscr{X}_{\mbox{\scriptsize o}}}
\newcommand{\mbbE}{\mathbb{E}}  
\newcommand{\mbbR}{\mathbb{R}}  

\newcommand{\Nc}{\mathcal{N}}

\newcommand{\Bbf}{{\bf {B}}}
\newcommand{\Hbf}{{\bf {H}}}

\newcommand{\Ibf}{{\bf {I}}}

\newcommand{\Ubf}{{\bf {U}}}
\newcommand{\Wbf}{{\bf {W}}}
\newcommand{\Vbf}{{\bf {V}}}

\newcommand{\abf}{{\bf {a}}}
\newcommand{\bbf}{{\bf {b}}}

\newcommand{\ebf}{{\bf {e}}}

\newcommand{\rbf}{{\bf {r}}}

\newcommand{\pbf}{{\bf {p}}}

\newcommand{\sbf}{{\bf {s}}}

\newcommand{\vbf}{{\bf {v}}}

\newcommand{\xbf}{{\bf {x}}}
\newcommand{\ybf}{{\bf {y}}}
\newcommand{\zbf}{{\bf {z}}}

\newcommand{\oscr}{\text{\scriptsize o}}
\newcommand{\cscr}{\text{\scriptsize c}}

\newcommand{\Ascr}{\mbox{\scriptsize A}}

\newcommand{\lisp}{\vspace{5pt}}

\def\QEDclosed{\mbox{\rule[0pt]{1.3ex}{1.3ex}}}
\def\QED{\QEDclosed}
\def\endproof{\hspace*{\fill}~\QED\par\endtrivlist\unskip}

\newcommand{\bitem}{\begin{itemize}}
\newcommand{\eitem}{\end{itemize}}
\newcommand{\bearr}{\begin{equation*}\begin{array}}
\newcommand{\eearr}{\end{array}\end{equation*}}
\newcommand{\beq}{\begin{equation*}}
\newcommand{\eeq}{\end{equation*}}
\newcommand{\bea}{\begin{eqnarray}}
\newcommand{\eea}{\end{eqnarray}}

\begin{document}

\title{Subspace Methods for Data Attack on State Estimation: A Data Driven Approach\thanks{
J. Kim, L. Tong, and R. J. Thomas are with the School of Electrical and Computer Engineering, Cornell University, Ithaca, NY 14853, USA. Email: {\tt jk752@cornell.edu, ltong@ece.cornell.edu, rjt1@cornell.edu}.  Part of this work was presented at the Asilomar Conference on Signals, Systems, and Computers, Pacific Grove, CA, november, 2013.
}}

\author{Jinsub Kim,
        Lang Tong,~\IEEEmembership{Fellow,~IEEE},
        and~Robert J. Thomas,~\IEEEmembership{Life Fellow,~IEEE}}%


\maketitle{\let\thefootnote\relax\footnotetext{
EDICS: SAM-APPL (Applications of sensor \& array multichannel processing), OTH-CPSY (Cyber-physical systems), SPE-DETC (Detection and estimation in power grid), SPE-PS (Privacy and security in power grid), and
SSP-IDEN (System identification).

This work is supported in part by the National Science Foundation under Grant CNS-1135844 and the Army Research Office under Grant W911NF1010419.}}

\begin{abstract}
Data attacks on state estimation modify part of system measurements such that the tempered measurements cause incorrect system state estimates.  Attack techniques proposed in the literature often require detailed knowledge of system parameters. Such information is difficult to acquire in practice.  The subspace methods presented in this paper, on the other hand, learn the system operating subspace from measurements and launch attacks accordingly.  Conditions for the existence of an unobservable subspace attack are obtained under the full and partial measurement models.  Using the estimated system subspace, two attack strategies are presented. The first strategy aims to affect the system state directly by hiding the attack vector in the system subspace.  The second strategy misleads the bad data detection mechanism so that data not under attack are removed.  Performance of these attacks are evaluated using the IEEE 14-bus network and the IEEE 118-bus network. 

\end{abstract}


\section{Introduction}

A cyber physical system (CPS) \cite{Lee2008UCB_TR} is a collection of physical devices networked by a cyber infrastructure with integrated sensing, communications, and control.  
A defining feature of CPS is coordinated operations based on data collected from sensors deployed throughout the system.  
Major examples of CPS include power grids, intelligent transportation systems, and networked robotics.

An essential signal processing component of many CPSs is real-time state estimation based on sensor measurements \cite{Huang2012:SPMag}.  The state estimate provides a CPS with the real-time monitoring and control capability.  
For instance, the state estimate of a power grid facilitates real-time economic dispatch, contingency analysis, and computation of real-time electricity price \cite{Huang2012:SPMag}.   

The dependency of CPS on data communications makes it vulnerable to cyber attacks where an adversary may break into the network, collect unauthorized information, and intercept and alter sensor data.   Because measurements are collected over a wide geographical area by distributed data acquisition systems, sometimes through wireless links, communications networks that support modern CPSs have numerous points of vulnerabilities \cite{Cardenas2009DHSWorkshop, Hull:2012PEMag}.  For critical infrastructures such as a power grid, a well planned coordinated attack may lead to a cascading failure and a regional blackout \cite{INL2011}. 


To assess vulnerability of CPS to possible cyber attacks, it is important to study potential attack mechanisms.  In this paper, we consider an adversary who can modify certain sensor data such that the corrupted data will mislead the CPS control with a wrong state estimate.  We refer to such a data attack on state estimation as a \emph{state attack}.  
A major challenge of state attack is to avoid being detected and identified by the fusion center.

In the literature, successful state attacks on a CPS, in particular a power grid, have been reported.  Liu, Ning, and Reiter \cite{Liu:2009CCS} presented the first state attack strategy, where an adversary replaces part of ``normal'' sensor data with ``malicious data.'' 
They showed that if an adversary can control a sufficiently large number of sensor data, it can perturb the state estimate by an arbitrary degree while avoiding detection at the control center.  Subsequent works along this line uncovered numerous attack and protection mechanisms \cite{Bobba&etal:10SCS, Kosut11, Kim&Poor:2011TSG, Bi&Zhang:2011Globecom, GianiEtal:2011SGC, KimTong:2013SGC, EsmalifalakEtal:2011SGC, Rahman:2012Globecom}.

Most proposed attack schemes require considerably detailed system information.  In particular, the network topology and physical system parameters are often required to construct attacks.  Although such information may be obtained by penetrating the control center, security measures can make it difficult in practice to access such information.

\subsection{Summary of contributions}

We consider the problem of data-driven attacks on state estimation, assuming that the adversary is capable of monitoring  a subset of system measurements without detailed knowledge of the network topology and system parameters.  The key idea in the proposed approach is to exploit the subspace structure of the measurements, in the same spirit of subspace techniques in array processing \cite{Stoica1989TASSP}, beamforming \cite{Pezeshki2008TSP}, and system identification \cite{Viberg1995Automatica}. 

The main contribution of this paper is the development of subspace techniques for state attack.  To this end, we present two techniques with different characteristics.  First, we show a construction of an unobservable attack based on the estimated subspace structure of measurements. We show further that, in constructing the attack, under certain conditions, monitoring only partial measurements may be sufficient.  In particular, we present a graph theoretic condition for the existence of an unobservable attack under the partial measurement model. 

The second subspace-based attack exploits the bad data detection and removal mechanisms. In particular, the attack purposely triggers the bad data detection, but it is designed to mislead the fusion center to remove data that are not tempered by the adversary while keeping some of the falsified data.    After such data removal, although the remaining data appear to be consistent with the system model, the resulting state estimate may have an arbitrarily large error.  We refer to this type of attack as \emph{data framing attack} in the sense that valid data are ``framed'' by the adversary and removed incorrectly by the fusion center.

To demonstrate the effectiveness of these attacks, we consider the problem of state estimation in a power system as a practical example of CPS.  To this end, we consider the IEEE 14-bus network and the IEEE 118-bus network\cite{IEEEParameter}.

An additional complexity of the power system is that the system observation is a nonlinear function of the system state. This raises the issue of whether attacks constructed from a linear model is effective in a nonlinear system.  While we do not have theoretical guarantees, simulation results show that the subspace-based data attacks perform well in the presence of nonlinearity in system equations.

\subsection{Related work and organization}

This paper extends some of the key results on state attacks that assume that the system parameters and the network topology are known to the attacker.   We describe below some of the relevant techniques. 

There is a substantial literature on state attacks when the system parameter and the network topology are known. 
Liu, Ning, and Reiter \cite{Liu:2009CCS} first introduced an \emph{unobservable attack} on power system state estimation, which can perturb the state estimate without being detected by the bad data detector at the fusion center.  
Following their seminal work, the link between feasibility of an unobservable attack and power system observability was made in  \cite{Kosut11,Bobba&etal:10SCS}.  Consequently, classical power system observability conditions \cite{Krumpholz&Clements&Davis:80PAS} can be modified to check feasibility of unobservable attacks and used to develop countermeasures based on sensor data authentication \cite{Bobba&etal:10SCS, Kosut11, Kim&Poor:2011TSG, Bi&Zhang:2011Globecom, Giani2013:TSG, KimTong:2013SGC, KimTong:13JSAC}.   
To assess the grid vulnerability against data attacks, the minimum number of adversary-controlled sensors necessary for an unobservable attack was suggested as the \emph{security index} of the grid \cite{Sandberg&Teixerira&Johansson:10SCS, Kosut11}.  
The data framing attack, when the system parameters are known, was first proposed in \cite{KimTongThomas:2013ArXiv} to circumvent the fundamental limit posed by the security index. 

There is limited work on state attacks without system information or with partial system information.  The use of independent component analysis in \cite{EsmalifalakEtal:2011SGC} is the most relevant. The authors of \cite{EsmalifalakEtal:2011SGC} proposed to identify a mixing matrix from which to construct an unobservable attack.  However, such techniques require that loads are statistically independent and non-Gaussian, and the techniques need full sensor observations. 
Generating unobservable attacks using partial parameter information was considered in \cite{Rahman:2012Globecom}.  The authors in \cite{Rahman:2012Globecom} showed that an adversary knowing impedance of transmission lines in a cutset of the network topology can construct an unobservable attack.  However, how an adversary can learn local parameters is nontrivial.  
In contrast to the aforementioned approaches, our method requires no system parameter information, and it can be launched with only partial sensor observations.
 
Attacks were also studied in the framework of a general dynamic CPS, under the assumption of an omniscient adversary.  
For instance, an attack on a linear control system equipped with a linear-quadratic-Gaussian controller was studied in \cite{MoSinopoli2010:CPSWEEK}.   Detectability and identifiability of attacks on general CPS operations was characterized in \cite{Pasqualetti2013:TAC}.  The model considered in these papers is more general than the static model studied here.  However, their assumption of an adversary with complete system information is stronger than that in the present work.

The rest of this paper is organized as follows.
Section~\ref{sec:math_model} introduces the measurement model, the mathematical model of state estimation and bad data processing, and the attack model.
Section~\ref{sec:unobservable_attack} presents the subspace methods of unobservable attack, and  
Section~\ref{sec:framing_attack} presents the subspace methods of data framing attack.
In Section~\ref{sec:numerical_result}, the results from simulations with benchmark power grids are presented.
Finally, Section~\ref{sec:conclusion_frame} provides concluding remarks.

\section{Mathematical models}\label{sec:math_model}

\subsection{Notations}

An upper case boldface letter (\eg, $\Hbf$) denotes a matrix, a lower case boldface letter (\eg, $\xbf$) denotes a vector, and a script letter (\eg, $\Amsc$, $\Smsc$) denotes a set.  The entry of $\Hbf$ at the $i$th row and the $j$th column is denoted by $\Hbf_{ij}$, and the $i$th entry of $\xbf$ is denoted by $x_{i}$. In addition, $\Rmsc(\Hbf)$ and $\Nmsc(\Hbf)$ denote the column space and the null space of $\Hbf$ respectively.  And, $\Ibf$ denotes an identity matrix with an appropriate size.  

\subsection{Measurement model}\label{subsec:measurement_model}

The \emph{system state} of a CPS is defined as a vector of variables that characterize the current operating condition of the CPS.   
We assume centralized state estimation at the fusion center.  For real-time estimation of the system state $\xbf\in\mbbR^{n}$, the fusion center collects measurements from sensors deployed throughout the system.   
Generally, the sensor measurements are related to the system state $\xbf$ in a nonlinear fashion, and the relation can be described by the nonlinear measurement model (\eg, the AC model for a power grid\cite{Abur&Exposito:book}):
\begin{equation}\label{eq:AC_baddata}
\zbf = h(\xbf) + \ebf,
\end{equation}
where $\zbf\in\mbbR^{m}$ is the measurement vector, $h(\cdot)$ is the measurement function, and $\ebf$ is the Gaussian measurement noise.

If some sensors malfunction or an adversary injects malicious data, the fusion center observes biased measurements,
\begin{equation}
\bar{\zbf} = h(\xbf) + \ebf + \abf,
\end{equation}
 where $\abf$ represents a deterministic bias.  In such a case, the data are said to be \emph{bad}, and the biased sensor entries are referred to as \emph{bad data entries}.  The bad data vector is typically sparse, and its support is unknown to the fusion center.  If $\abf$ is injected by an adversary, $\abf$ is constrained by its support. 

In analyzing the attack effect on state estimation, we adopt a linearization of (\ref{eq:AC_baddata}) around a nominal state $\xbf_{0}$:
\begin{equation}\label{eq:DC_baddata_1}
\zbf = h(\xbf_{0}) +  \Hbf(\xbf - \xbf_{0}) + \ebf,
\end{equation}
where $\Hbf\in\mbbR^{m\times n}$ is the measurement matrix that relates the system state to the measurement vector, and $\ebf$ is the Gaussian measurement noise with a covariance matrix $\sigma^{2}\Ibf$.   
Without loss of generality, we assume that both $h(\xbf_{0})$ and $\xbf_{0}$ are zero vectors\footnote{For general cases, we can simply treat $\zbf_{1} \triangleq \zbf-h(\xbf_{0})$ and $\xbf_{1}\triangleq\xbf - \xbf_{0}$ as the measurement vector and the state vector and work with $\zbf_1 = \Hbf\xbf_1 + \ebf$.} and employ the following model: 
\begin{equation}\label{eq:DC_baddata}
\zbf =  \Hbf\xbf + \ebf.
\end{equation}

A system is said to be \emph{observable} if the measurement matrix $\Hbf$ has full column rank (\ie, $\xbf$ can be uniquely determined from $\Hbf\xbf$.)  System observability is essential for state estimation.  In practice, sensors should be placed in the network to satisfy observability.  Hence, we assume that the CPS of interest is observable, \ie, $\Hbf$ has full column rank.

In practice, the nonlinear system and the nonlinear iterative state estimation techniques have a certain mitigating effect on attacks designed based on a linear model\cite{Jia2014:TPS}.  It is therefore important to validate performance of an attack strategy based on the nonlinear model (\ref{eq:AC_baddata}) using a nonlinear state estimator.  Note that, while our attacks are constructed based on (\ref{eq:DC_baddata}), our numerical experiments validate their performance using the original nonlinear system (\ref{eq:AC_baddata}) with a nonlinear state estimator.

\subsection{State estimation and bad data processing}\label{subsec:SE}

This section introduces a popular approach to state estimation and bad data processing \cite{Handschin&Schweppe&Kohlas&Feichter:75TPAS, Abur&Exposito:book}, which we assume to be employed by the fusion center.   The specific approach is a widely used standard implementation in the power grid where the number of states is in the order of 10,000, and the estimates are made every few minutes.  

\begin{figure}[t!]
\centering
\psfrag{z}[c]{ $\zbf$ }
\psfrag{zb}[c]{ $\bar{\zbf}$ }
\psfrag{g}[c]{ $\Gmsc$}
\psfrag{gb}[c]{ ${\Gmsc}$}
\psfrag{x}[c]{ $\hat{\xbf}$ }
\psfrag{o}[l]{ $(\hat{\xbf},\Gmsc)$}
\psfrag{ob}[l]{$\hat{\xbf}$}
\psfrag{n}[c]{ $\sbf$}
\includegraphics[width=.4\textwidth]{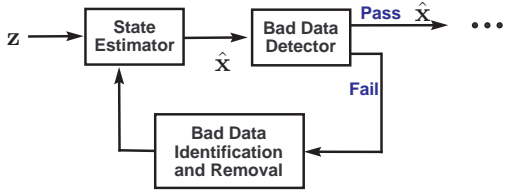}
\caption{State estimation and bad data processing}
\label{fig:se_attack_analog}
\end{figure}

Fig.~\ref{fig:se_attack_analog} illustrates an iterative scheme for obtaining an estimate $\hat{\xbf}$ of the system state, which consists of three functional blocks: state estimation, bad data detection, and bad data identification.

The assumed state estimator is based on the maximum likelihood principle and is implemented in a recursive manner.  
Iterations begin with the initial measurement vector $\zbf^{(1)} \triangleq \zbf$ and the initial measurement function $h^{(1)} \triangleq h$ where the superscript denotes the index for the current iteration.  

In the $k$th iteration, state estimation uses $(\zbf^{(k)}, h^{(k)})$ as an input and calculates the least squares (LS) estimate of the system state and the corresponding residue vector:
\begin{equation}\label{eq:WLS_AC}
\begin{array}{l}
\hat{\xbf}^{(k)}\triangleq \arg\min_{\xbf}\dfrac{1}{\sigma^{2}}\|\zbf^{(k)} - h^{(k)}(\xbf)\|_{2}^{2},\\[7pt]
\rbf^{(k)}\triangleq \zbf^{(k)} - h^{(k)}(\hat{\xbf}^{(k)}),
\end{array}
\end{equation}
where $\|\cdot\|_{2}$ denotes $l_{2}$ norm. 
In practice, the above nonlinear LS estimate can be obtained by iteration of a linearized LS estimation using Newton-Raphson or quasi-Newton methods\cite{Abur&Exposito:book}.


Bad data detection employs the $J(\hat{\xbf})$-test\cite{Handschin&Schweppe&Kohlas&Feichter:75TPAS, Abur&Exposito:book}:
\begin{equation}\label{eq:BDD}
\left\{\begin{array}{ll}
\text{bad data} & \text{if $\dfrac{1}{\sigma^{2}}\|\rbf^{(k)}\|_{2}^{2} > \tau^{(k)}$;}\\[7pt]
\text{good data} & \text{if $\dfrac{1}{\sigma^{2}}\|\rbf^{(k)}\|_{2}^{2} \leq \tau^{(k)}$}
\end{array}
\right.
\end{equation}
where $\tau^{(k)}$ is a predetermined threshold.  
The $J(\hat{\xbf})$-test is widely used due to its simplicity and the fact that the test statistic has a $\chi^{2}$ distribution if the data are good \cite{Handschin&Schweppe&Kohlas&Feichter:75TPAS}.  The latter fact is used to set the threshold $\tau^{(k)}$ for a given false alarm constraint.  

If the bad data detector (\ref{eq:BDD}) declares that the data are good, the algorithm returns the state estimate $\hat{\xbf}^{(k)}$ and \emph{terminates}.  
However, if the bad data detector declares that the data are bad, bad data identification is invoked to identify and remove \emph{one} bad data entry from the measurement vector.

A widely used criterion for identifying a bad data entry is the normalized residue\cite{Handschin&Schweppe&Kohlas&Feichter:75TPAS, Abur&Exposito:book}:  
 each $r^{(k)}_{i}$ is divided by its standard deviation under the hypothesis that $\zbf^{(k)}$ contains no bad data.  Therefore, each normalized residue approximately follows the standard normal distribution if $\zbf^{(k)}$ contains no bad data. 
Specifically, 
\begin{equation}\label{eq:normalize}
\tilde{\rbf}^{(k)} \triangleq \bm{\Omega}^{(k)}\rbf^{(k)},
\end{equation}
where $\bm{\Omega}^{(k)}$ is a diagonal matrix with 
\begin{equation}\label{eq:norm_matrix}
\bm{\Omega}^{(k)}_{ii} \triangleq\left\{\begin{array}{ll}
 0 & \hspace{-5pt}\begin{array}{l}\text{if removing $i$ makes}\\ \text{the system unobservable\footnotemark;}\end{array}\\[4pt]
\dfrac{1}{\sqrt{\sigma^{2}\Wbf^{(k)}_{ii}}} & \text{otherwise;}
\end{array}\right.
\end{equation}
\footnotetext{If removing the sensor $i$ makes the system unobservable, its residue is always equal to zero\cite{Abur&Exposito:book}, and the corresponding diagonal entry of $\Wbf^{(k)}$ is zero.  For such a sensor, the normalizing factor is $0$ such that its normalized residue is equal to $0$.}
and $\Wbf^{(k)}$ is defined as
\begin{equation}
\Ibf - \Hbf^{(k)}((\Hbf^{(k)})^{T}\Hbf^{(k)})^{-1}{(\Hbf^{(k)})}^{T}
\end{equation}
 with $\Hbf^{(k)}$ denoting the Jacobian of $h^{(k)}$ at $\hat{\xbf}^{(k)}$ (see Appendix of \cite{Handschin&Schweppe&Kohlas&Feichter:75TPAS} for details.)

Once the normalized residue $\tilde{\rbf}^{(k)}$ is calculated, the sensor with the largest $|\tilde{r}^{(k)}_{i}|$ is identified as a bad sensor.  The row of $\zbf^{(k)}$ and the row of $h^{(k)}$ that correspond to the bad sensor are removed, and the updated measurement vector $\zbf^{(k+1)}$ and measurement function $h^{(k+1)}$ are used as the inputs for the next iteration.


Using the linearized model (\ref{eq:DC_baddata}), every step is the same as using the nonlinear model, except that the nonlinear measurement function $h^{(k)}(\xbf)$ is replaced with the linear function $\Hbf^{(k)}\xbf$ (so, the Jacobian is the same everywhere.) 
Note that the LS state estimate (\ref{eq:WLS_AC}) is replaced with a simple linear LS solution:
\begin{equation}\label{eq:WLS_DC}
\hat{\xbf}^{(k)} = ((\Hbf^{(k)})^{T}\Hbf^{(k)})^{-1}{(\Hbf^{(k)})}^{T}\zbf^{(k)},
\end{equation}
and thus
\begin{equation}\label{eq:r_DC}
\rbf^{(k)} = \zbf^{(k)} - \Hbf^{(k)}\hat{\xbf}^{(k)} = \Wbf^{(k)}\zbf^{(k)}.
\end{equation}

\subsection{Adversary model}

An adversary is assumed to be capable of modifying the data from a subset of sensors $\SmscA$, referred to as \emph{adversary sensors}.  The fusion center observes corrupted measurements $\bar{\zbf}$ instead of the real measurements $\zbf$.  The adversarial modification is mathematically modeled by:
\begin{equation}\label{eq:attack_baddata}
\bar{\zbf} = \zbf + \abf,~~~\abf\in\Amsc,
\end{equation}
where $\abf$ is an attack vector, and $\Amsc$ is the set of feasible attack vectors defined as 
\begin{equation}
\Amsc\triangleq\{\abf\in\mbbR^{m}:~a_{i} = 0,~\forall i\notin\SmscA\}.
\end{equation}

Liu, Ning, and Reiter \cite{Liu:2009CCS} presented an \emph{unobservable attack}, which is a powerful attack mechanism capable of perturbing the state estimate without being detected.
An unobservable attack can be formally defined as follows.

\lisp
\begin{definition}
Given a measurement vector $\zbf$ corresponding to a state $\xbf$, \ie, $\zbf=\Hbf\xbf+\ebf$, 
a state attack $\abf \in \Amsc$ is \emph{unobservable} if there exists a state $\bar{\xbf} \neq \xbf$ such that $\zbf+\abf = \Hbf\bar{\xbf}+\ebf$.
\end{definition}
\lisp

The following Lemma shows the algebraic property of the attack; it follows immediately from the definition.

\lisp
\begin{lemma}
A state attack is unobservable if and only if $\abf\neq\textbf{0}$, and $\abf \in \Rmsc(\Hbf) \cap \Amsc$.  Furthermore, if $\abf$ is unobservable, so is $\gamma\cdot \abf$ for any nonzero $\gamma\in\mbbR$, and $\|\xbf-\bar{\xbf}\|_{2} \rightarrow \infty$ as $\gamma \rightarrow \infty$.
\end{lemma}
\lisp

The feasibility of an unobservable attack is closely related to the concept of system observability. 
In particular, the following connection was found in \cite{Kosut11}.

\lisp
\begin{theorem}[\hspace{-0.2pt}\cite{Kosut11}]\label{thm:feasibility_unobservable}
An unobservable attack is feasible if and only if removing the adversary sensors makes the grid unobservable (\ie, the measurement matrix does not have full column rank.)
\end{theorem}
\lisp

\begin{proof}
See Appendix~\ref{app:Kosut}.  
\end{proof}
\lisp
  
\section{Subspace methods for unobservable attack}\label{sec:unobservable_attack}
Most existing works on an unobservable attack assumed that an adversary knows the measurement matrix $\Hbf$. 
In contrast, this section presents a design of an unobservable attack based on the system measurement subspace, without knowledge of $\Hbf$.   Employing the linearized measurement model (\ref{eq:DC_baddata}), we will present the conditions under which an unobservable attack can be constructed based on the subspace information.  We also demonstrate a condition that guarantees the design of an unobservable attack based on partial sensor measurements; for an attack on a power grid, this condition is characterized as a graph condition on the network topology.  
\subsection{Feasibility of an unobservable attack}\label{subsec:unobservable_subspace}

Note that designing an unobservable attack is equivalent to finding a nonzero vector in $\Rmsc(\Hbf)$ satisfying the sparsity pattern defined by $\Amsc$.
Therefore, an unobservable attack, if feasible, can be launched by using a basis matrix $\Ubf\in\mbbR^{m\times n}$ of $\Rmsc(\Hbf)$ without knowing $\Hbf$, as stated in the following theorem.   
Formally, we refer to $\Rmsc(\Hbf)$ as the \emph{measurement subspace} because it is the subspace of all possible noiseless measurements. 

\lisp
\begin{theorem}\label{thm:unobservable_subspace}
Let $\Ubf$ be any basis matrix of $\Rmsc(\Hbf)$ and $\bar{\Ubf}$ a submatrix of $\Ubf$ obtained by removing the rows corresponding to the adversary sensors.  Then, the following are true:
\begin{enumerate}
\item An unobservable attack is feasible if and only if $\bar{\Ubf}$ does not have full column rank.
\item When feasible, an unobservable attack can be constructed using $\Ubf$: for a nonzero vector $\vbf\in\Nmsc(\bar{\Ubf})$, $\abf\triangleq\Ubf\vbf$ is an unobservable attack vector.
\end{enumerate}
\end{theorem}
\lisp

\begin{proof}
See Appendix~\ref{app:unobservable_subspace}.  
\end{proof}
\lisp

Note that in constructing the unobservable attack vector $\Ubf\vbf$, all that is necessary is a basis matrix $\Ubf$ of $\Rmsc(\Hbf)$.

\subsection{Unobservable attack with partial measurements}\label{subsec:unobservable_partial}
In this section, we show that an unobservable attack can be constructed using the subspace information of \emph{partial} sensor measurements.  
To formally state the result, we need the notion of a {critical set} of sensors \cite{Abur&Exposito:book} and {partial observability} defined as follows.

\lisp
\begin{definition}\label{def:parital_observability}
A set of sensors is called a \emph{critical set} if removing the set of sensors from the system renders the system unobservable while removing any strict subset of it does not. 
Let $\Smsc$ and $\Xmsc$ denote a subset of sensors and a subset of state variables respectively.   The state variables in $\Xmsc$ are said to be \emph{observable with respect to} $\Smsc$ if the state variables in $\Xmsc$ can be uniquely determined based on measurements from $\Smsc$\footnote{In other words, every element of $\Nmsc(\Hbf_{\text{s}})$ has zero entries for the rows corresponding to the state variables in $\Xmsc$, where $\Hbf_{\text{s}}\in\mbbR^{|\Smsc|\times n}$ is the submatrix of $\Hbf$ obtained by retaining only the rows corresponding to the sensors in $\Smsc$.}.  When the state variables in $\Xmsc$ are observable with respect to $\Smsc$, a subset $\Cmsc$ of $\Smsc$ is a \emph{critical set with respect to} $(\Smsc, \Xmsc)$ if removing $\Cmsc$ from $\Smsc$ makes the state variables in $\Xmsc$ no longer observable with respect to $\Smsc$ while removing a strict subset of $\Cmsc$ from $\Smsc$ does not.  
\end{definition}
\lisp

Consider a subset of sensors $\Smsco$.  Let $\Xmsco$ denote the set of state variables whose values affect measurements from the sensors in $\Smsco$ (\ie, the $|\Smsco|$ by $n$ submatrix $\Hbf_\oscr$ of $\Hbf$, consisting of the rows corresponding to the sensors in $\Smsco$, has nonzero columns exactly at the columns corresponding to the state variables in $\Xmsco$.) 

The following theorem provides the conditions under which an unobservable attack can be constructed based on the subspace information of measurements from $\Smsco$.  The conditions roughly mean that (i) based on measurements from $\Smsco$, one can uniquely identify the relevant state variables (\ie, the variables in $\Xmsco$,) 
and (ii) $\Smsco$ contains a set of sensors, which, if controlled by an adversary, is sufficient for launching an unobservable attack and is also critical with respect to $(\Smsco, \Xmsco)$.   

\vspace{5pt}
\begin{theorem}\label{thm:unobservable_partial}
Suppose that 
\begin{enumerate}
\item the state variables in $\Xmsco$ are observable with respect to $\Smsco$,
\item $\Cmsc\subset\Smsco$ is a critical set with respect to $(\Smsco, \Xmsco)$, and
\item removing $\Cmsc$ makes the system unobservable.
\end{enumerate}
Let $\Hbf_{\oscr}\in\mbbR^{|\Smsc_{\text{\tiny o}}|\times n}$ denote the submatrix of $\Hbf$ obtained by retaining only the rows corresponding to the sensors in $\Smsco$.  Then, the following are true:
\begin{enumerate}
\item Let $\Amsc_{\oscr}$ denote the set of vectors in $\Rmsc(\Hbf_{\oscr})$ such that $\bbf\in\Rmsc(\Hbf_{\oscr})$ is in $\Amsc_{\oscr}$ if and only if the rows of $\bbf$ corresponding to the sensors in $\Smsco\setminus\Cmsc$ are equal to zero.  Then, the dimension of $\Amsc_{\oscr}$ is one. 
\item For an arbitrary \emph{nonzero} $\abf_{\oscr}\in\Amsc_{\oscr}$, the attack that modifies the sensor data from $\Cmsc$ by adding the corresponding entries in $\abf_{\oscr}$ to the real data is unobservable.  
\end{enumerate}
\end{theorem}
\vspace{3pt}
\begin{proof}
See Appendix~\ref{app:unobservable_partial}.
\end{proof}
\vspace{5pt}

Note that $\Amsc_{\oscr}$ in Theorem~\ref{thm:unobservable_partial} can be fully characterized based on a basis matrix of $\Rmsc(\Hbf_{\oscr})$.  
 The following corollary provides the detail of how an attack can be constructed from a basis matrix of $\Rmsc(\Hbf_{\oscr})$.

\vspace{5pt}
\begin{corollary}\label{cor:local_unobservable}
Suppose that the conditions 1), 2), and 3) of Theorem~\ref{thm:unobservable_partial} hold.  
Let $\Ubf_{\oscr}\in\mbbR^{|\Smsc_{\text{\tiny o}}|\times |\Xmsc_{\text{\tiny o}}|}$ denote a basis matrix of $\Rmsc(\Hbf_{\oscr})$ and $\bar{\Ubf}_{\oscr}$ denote a submatrix of $\Ubf_{\oscr}$ obtained by removing the rows corresponding to the sensors in $\Cmsc$.  Then, the following are true:
\begin{enumerate}
\item The dimension of $\Nmsc(\bar{\Ubf}_{\oscr})$ is one.
\item For any nonzero vector $\vbf\in\Nmsc(\bar{\Ubf}_{\oscr})$, the attack that modifies the sensor data from $\Cmsc$ by adding the corresponding entries in $\Ubf_{\oscr}\vbf$ to the real data is unobservable.  
\end{enumerate}
\end{corollary}
\vspace{5pt}

The three conditions of Theorem~\ref{thm:unobservable_partial} are all related to system observability or partial observability.  In case of a power grid, system observability and partial observability can be checked based on \emph{partial} information about the grid topology and sensor locations.  
In particular, the graph-theoretical observability criterion in \cite{Krumpholz&Clements&Davis:80PAS} can be employed.  
  
A power grid is a network of buses connected by transmission lines.  The \emph{topology} of a grid is naturally defined as an undirected graph $\Gmsc = (\Vmsc, \Emsc)$ where $\Vmsc$ is the set of buses, and $\Emsc$ is the set of connected transmission lines: $\{i,j\}$ is in $\Emsc$ if and only if there exists a connected transmission line between bus $i$ and bus $j$.  
We consider two types of legacy sensors: line flow sensors and bus injection sensors.  A line flow sensor located on a line $\{i,j\}$ measures the power flowing through the line either from bus $i$ to bus $j$ or from bus $j$ to bus $i$.  A bus injection sensor on bus $i$ measures the total power injected into the network at bus $i$ (see Appendix~\ref{app:example} for the details of the sensor measurements.)  

\begin{figure}[t!]
\centering
\includegraphics[width=.35\textwidth]{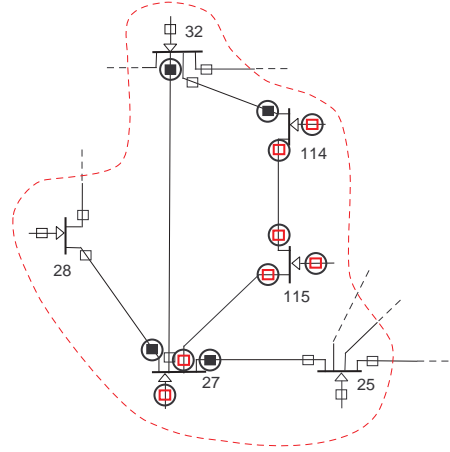}
\caption{A part of the IEEE 118-bus network: Rectangles represent the sensor locations. Every bus has an injection sensor, and every line has line flow sensors for both directions.}  
\label{fig:118bus_part}
\end{figure}

The following corollary presents the graph conditions that imply the conditions of Theorem~\ref{thm:unobservable_partial} for an attack on a power grid state estimation.  
 Appendix~\ref{app:example} provides the details of the graph-theoretical observability criterion in \cite{Krumpholz&Clements&Davis:80PAS}, which directly results in the following corollary from Theorem~\ref{thm:unobservable_partial}.  To state the corollary, we need to introduce the concept of a \emph{reduced power network}.  Given a subset $\Smsco$ of sensors, the reduced network consists of the sensors in $\Smsco$ and the topology $\bar{\Gmsc} = (\bar{\Vmsc}, \bar{\Emsc})$, where $\{i,j\}$ is in $\bar{\Emsc}$ if and only if a line flow sensor on $\{i,j\}$ is in $\Smsco$, or an injection sensor at bus $i$ or bus $j$ is in $\Smsco$, and $\bar{\Vmsc}$ consists of all the endpoints of the lines in $\bar{\Emsc}$.  
For instance, in the IEEE 118-bus network,  Fig.~\ref{fig:118bus_part} describes a reduced network for $\Smsco$ consisting of the circled sensors.  In this example, the vertices and edges inside the dashed boundary form $\bar{\Gmsc}$.  

\lisp
\begin{corollary}\label{cor:power_grid}
Let $\Smsco$ be a subset of sensors, $\bar{\Gmsc} = (\bar{\Vmsc}, \bar{\Emsc})$ the topology of the reduced network for $\Smsco$, and $\Cmsc$ a subset of $\Smsco$.  Suppose that 
\begin{enumerate}
\item There exists a cut of the grid topology $\Gmsc$ such that $\Cmsc$ consists of all line flow sensors on the cutset lines and all injection sensors on the endpoints of the cutset lines.
\item For every sensor $s$ in $\Cmsc$, there exists a way to assign each injection sensor in $(\Smsco\setminus\Cmsc)\cup\{s\}$ to a line incident to the bus where the sensor is located\footnote{In other words, for an injection sensor located at bus $i$, we assign the injection sensor to one of the lines that are incident to bus $i$.  We do this for each injection sensor in $(\Smsco\setminus\Cmsc)\cup\{s\}$.} such that there exists a spanning tree of $\bar{\Gmsc}$ with at least one sensor in $(\Smsco\setminus\Cmsc)\cup\{s\}$ on every edge of the tree (either a line flow or an assigned injection sensor.)
\end{enumerate}
Then, the conditions of Theorem~\ref{thm:unobservable_partial} hold, and thus the statements in Theorem~\ref{thm:unobservable_partial} and Corollary~\ref{cor:local_unobservable} hold.
\end{corollary}
\lisp

Note that the conditions of Corollary~\ref{cor:power_grid} are related to the topology and the sensor locations in the reduced network.  Therefore, an adversary can exploit partial information about the topology and sensor locations to find an attack setting that enables an unobservable attack with partial sensor observations.  For instance, it can be easily checked that the example in Fig.~\ref{fig:118bus_part} with $\Cmsc$ consisting of the circled empty-rectangle sensors satisfies the conditions.  In particular, the first condition is satisfied with the cut that isolates bus 115 from the rest of the network.  

\subsection{Subspace attack algorithm}\label{subsec:unobservable_data}
All the information necessary for subspace attack methods is the subspace information of $\Rmsc(\Hbf)$ or $\Rmsc(\Hbf_{\oscr})$.   Subspace estimation based on measurement data has been actively studied in the signal processing literature (\eg, \cite{Srivastava2000TSP, Smith2005TSP}), and thus subspace methods naturally lead to a data-driven algorithm for practical attack scenarios.  Our focus in this section is to demonstrate how (any) subspace estimator can be used to generate a data-driven attack.

One of the simplest yet effective ways of estimating a basis matrix is to use a sample covariance matrix.  
Let $\zbf_{1},\ldots,\,\zbf_{K}$ denote measurement vectors at $K$ different sampling instances:
\begin{equation}
\zbf_{i} = \Hbf\xbf_{i} + \ebf_{i},~~i=1,\ldots,\,K.
\end{equation}
For simplicity, suppose that the noise vectors $\ebf_{1},\ldots,\,\ebf_{K}$ are independent and identically distributed (i.i.d.), the state vectors $\xbf_{1},\ldots,\,\xbf_{K}$ are i.i.d. with a positive definite covariance matrix $\bm{\Sigma}_{\xbf}$, and the noise vectors and the state vectors are uncorrelated.  Then, the covariance matrix of $\zbf$ is
\begin{equation}
\bm{\Sigma}_\zbf\triangleq \mbbE\left[(\zbf_{1} - \mbbE[\zbf_{1}])(\zbf_{1} - \mbbE[\zbf_{1}])^{T}\right] = \Hbf\bm{\Sigma}_{\xbf}\Hbf^{T} + \sigma^{2}\Ibf.
\end{equation}
Note that $\Hbf\bm{\Sigma}_{\xbf}\Hbf^{T}$ has rank $n$.
Therefore, if $\Ubf\bm{\Lambda}\Vbf^{T}$ is a singular value decomposition (SVD) of $\bm{\Sigma}_\zbf$, the $n$ columns of $\Ubf$ that correspond to the $n$ largest singular values form a basis of $\Rmsc(\Hbf\bm{\Sigma}_{\xbf}\Hbf^{T})$.  Because $\Rmsc(\Hbf\bm{\Sigma}_{\xbf}\Hbf^{T})$ is equivalent to $\Rmsc(\Hbf)$, the same columns form a basis of $\Rmsc(\Hbf)$. 

Therefore, in practice, we can estimate a basis matrix of $\Rmsc(\Hbf)$ by applying SVD to the sample covariance matrix $\hat{\bm{\Sigma}}_{\zbf}$: 
\begin{equation}
\hat{\bm{\Sigma}}_{\zbf} \triangleq \dfrac{1}{K-1}\sum_{i=1}^{K} (\zbf_{i} - \underbar{\zbf}) (\zbf_{i} - \underbar{\zbf})^{T},
\end{equation}
where $\underbar{\zbf}$ denotes the sample mean.

Based on the above (or any other) subspace estimator and Theorem~\ref{thm:unobservable_subspace}, the data-driven attack with \emph{full} sensor observations operates as follows with the observations $\{\zbf_{1},\ldots,\,\zbf_{K}\}$ and the adversary sensor set $\SmscA$ as inputs: 
\begin{enumerate}
\item \textbf{Subspace estimation}: Based on $\{\zbf_{1},\ldots,\,\zbf_{K}\}$, calculate an estimate $\hat{\Ubf}\in\mbbR^{m\times n}$ of a basis matrix of $\Rmsc(\Hbf)$.
\item \textbf{Null space estimation}: Obtain $\hat{\Ubf}_{1}$ by removing the rows of $\hat{\Ubf}$ that correspond to the sensors in $\SmscA$.  Find an SVD of $\hat{\Ubf}_{1}$, $\hat{\Ubf}_{1} = \tilde{\Ubf}\tilde{\bm{\Lambda}}\tilde{\Vbf}^{T}$, and let $\vbf$ denote the column of $\tilde{\Vbf}$ that corresponds to the smallest singular value ($\vbf$ is an estimate of a nonzero element of $\Nmsc(\bar{\Ubf})$ in Theorem~\ref{thm:unobservable_subspace}.)
\item \textbf{Attack}: Modify the sensor data from $\SmscA$ by adding the corresponding entries of $\eta\cdot\hat{\Ubf}\vbf$ to them, where $\eta\in\mbbR$ is a scaling factor to adjust the degree of perturbation.
\end{enumerate}

The data-driven attack with \emph{partial} sensor observations can be constructed in the same manner based on Corollary~\ref{cor:local_unobservable}.  Specifically, the attack receives $(\Xmsco,\Smsco, \Cmsc)$ and $\{\tilde{\zbf}_{1},\ldots,\,\tilde{\zbf}_{K}\}$---the set of measurements from the sensors in $\Smsco$ at $K$ different time instances---as inputs and executes the following steps:
\begin{enumerate}
\item \textbf{Subspace estimation}: Based on $\{\tilde{\zbf}_{1},\ldots,\,\tilde{\zbf}_{K}\}$, calculate an estimate $\hat{\Ubf}_{\oscr}\in\mbbR^{|\Smsc_{\text{\tiny o}}|\times |\Xmsc_{\text{\tiny o}}|}$ of a basis matrix of $\Rmsc(\Hbf_{\oscr})$.

\item \textbf{Null space estimation}: Obtain $\hat{\Ubf}_{\cscr}$ by removing the rows of $\hat{\Ubf}_{\oscr}$ that correspond to the sensors in $\Cmsc$.  Find an SVD of $\hat{\Ubf}_{\cscr}$: $\hat{\Ubf}_{\cscr} = \tilde{\Ubf}\tilde{\bm{\Lambda}}\tilde{\Vbf}^{T}$.  Let $\vbf$ denote the column of $\tilde{\Vbf}$ that corresponds to the smallest singular value ($\vbf$ is an estimate of a nonzero element of $\Nmsc(\bar{\Ubf}_{\oscr})$ in Corollary~\ref{cor:local_unobservable}.) 

\item \textbf{Attack}: Modify the sensor data from $\Cmsc$ by adding the corresponding entries of $\eta\cdot\hat{\Ubf}_{\oscr}\vbf$ to them, where $\eta\in\mbbR$ is a scaling factor to adjust the degree of perturbation.
\end{enumerate}

\section{Subspace methods for data framing attack}\label{sec:framing_attack}

The idea of data framing attack based on full system parameter information was first presented in \cite{KimTongThomas:2013ArXiv}.  In this section, we demonstrate data-driven approaches of data framing attack by exploiting the subspace structure of sensor measurements.  

\subsection{Data framing attack}\label{subsec:prelim_framing}

A data framing attack aims to enable an adversary to perturb the state estimate by an arbitrary degree even when an unobservable attack with $\SmscA$ does not exist.  
To this end, a data framing attack frames some normally operating meters as sources of bad data such that their data will be removed.   
A critical parameter of data framing attack is the set of sensors to be framed, denoted by $\SmscF$.  
The framed sensor set $\SmscF$ is selected such that $\SmscF\cap\SmscA = \emptyset$, and if the sensors in $\SmscF$ are removed from the system, an unobservable attack with $\SmscA$ becomes feasible.   Under this selection rule, an adversary may design an attack that becomes unobservable once the sensor data from $\SmscF$ are removed by the bad data removal rule.  

To successfully make the data from $\SmscF$ removed, one can use an attack vector that maximizes the energy of the normalized residues at $\SmscF$ in the \emph{first} iteration of the bad data processing.  
Such an attack design does not necessarily guarantee that all data from $\SmscF$ will be identified as bad.  Nevertheless, this is a reasonable heuristic to circumvent the difficulty of analyzing attack effect on normalized residues in all iterations.  

To simplify notation, we drop the superscript that denotes the first iteration of bad data processing: all the quantities in this section are from the first iteration unless otherwise specified.  
The attack \emph{direction} that maximizes the energy of the normalized residues in the first iteration can be constructed by solving the following optimization\cite{KimTongThomas:2013ArXiv}:
\begin{equation}\label{eq:concept}
\begin{array}{ll}
\max_{\abf} & \mbbE\left[\sum_{i\in\Smsc_{\text{\tiny F}}}(\tilde{r}_{i})^{2}\right]\\
\text{subj.} & \|\abf\|_{2}^{2} = 1, ~~ \abf\in\Rmsc(\Hbf_{1})\cap\Amsc,
\end{array}
\end{equation}
where $\Hbf_{1}\in\mbbR^{m\times n}$ is a matrix obtained from $\Hbf$ by replacing the rows corresponding to the sensors in $\SmscF$ with zero row vectors.   
The constraint $\abf\in\Rmsc(\Hbf_{1})$ holds if and only if $\abf$ is unobservable after the framed sensor data are removed.  This constraint guarantees that once the data from $\SmscF$ are removed, the attack can have the same effect as an unobservable attack.  


The following theorem states that a solution to (\ref{eq:concept}) can be obtained without knowing $\Hbf$ if we know a basis matrix of $\Rmsc(\Hbf)$.  

\vspace{5pt}
\begin{theorem}\label{thm:subspace_framing}
An adversary knowing a basis matrix $\Ubf\in\mbbR^{m\times n}$ of $\Rmsc(\Hbf)$ can find a solution of (\ref{eq:concept}).  Specifically, a solution to the following quadratically constrained quadratic programming (QCQP) is also a solution to (\ref{eq:concept}), and vice versa:
\begin{equation}\label{eq:qcqp_2}
\begin{array}{ll}
\max_{\abf} & \|\Ibf_{\Smsc_{\text{\tiny F}}}\tilde{\bm{\Omega}} \tilde{\Wbf}\abf\|_{2}^{2}\\
\text{subj.} & \|\abf\|_{2}^{2} = 1, ~~ \abf\in\Rmsc(\Ubf_{1})\cap\Amsc,
\end{array}
\end{equation}
where $\Ibf_{\Smsc_{\text{\tiny F}}}\in\mbbR^{|\Smsc_{\text{\tiny F}}|\times m}$ is the row selection operator that retains only the rows corresponding to the sensors in $\SmscF$ out of $m$ rows, 
\begin{equation}
\tilde{\Wbf} \triangleq \Ibf - \Ubf(\Ubf^{T}\Ubf)^{-1}\Ubf^{T},
\end{equation}
$\tilde{\bm{\Omega}}\in\mbbR^{m\times m}$ is a diagonal matrix with 
\begin{equation}
\tilde{\bm{\Omega}}_{ii} = \left\{\begin{array}{ll}
1/\sqrt{\tilde{\Wbf}_{ii}} &\text{if $\tilde{\Wbf}_{ii}> 0$;}\\
0 & \text{if $\tilde{\Wbf}_{ii} = 0$,}
\end{array}
\right.
\end{equation} 
and $\Ubf_{1}\in\mbbR^{m\times n}$ is a matrix obtained from $\Ubf$ by replacing the rows corresponding to the sensors in $\SmscF$ with zero row vectors.
\end{theorem}
\vspace{5pt}

\begin{proof}
See Appendix~\ref{app:subspace_framing}.
\end{proof}
\vspace{5pt}

Note that addition of the attack vector $\abf$ changes the mean of the residue vector from $\textbf{0}$ to $\tilde{\Wbf}\abf$.   
And, $\Ibf_{\Smsc_{\text{\tiny F}}}\tilde{\bm{\Omega}} \tilde{\Wbf}\abf/\sigma$ is the resulting mean of the normalized residues of the data from $\SmscF$.

\subsection{Sufficiency of partial measurements}\label{subsec:subspace_framing_partial}

Similar to sufficiency of partial measurements for an unobservable attack (Theorem~\ref{thm:unobservable_partial}), 
data framing attack can also be launched based on subspace information of partial measurements, as stated formally in the following theorem.  Below, we use the notations defined in Section~\ref{subsec:unobservable_partial} for the partial measurement case.

\vspace{5pt}
\begin{theorem}\label{thm:subspace_partial_framing}
Suppose that the conditions 1), 2), and 3) of Theorem~\ref{thm:unobservable_partial} hold for $\Smsco$, $\Xmsco$, and $\Cmsc$.  
Let $\{\Cmsc_{1},\,\Cmsc_{2}\}$ denote an arbitrary partition of $\Cmsc$.
Let $\Hbf_{\Ascr}$ denote a submatrix of $\Hbf$ consisting of the rows corresponding to the sensors in $\Smsco\setminus\Cmsc_{2}$, $\Ubf_{\Ascr}\in\mbbR^{|\Smsc_{\text{\tiny o}}\setminus\Cmsc_{2}|\times |\Xmsc_{\text{\tiny o}}|}$ denote a basis matrix of $\Rmsc(\Hbf_{\Ascr})$, and $\bar{\Ubf}_{\Ascr}$ denote a submatrix of $\Ubf_{\Ascr}$ obtained by removing the rows corresponding to the sensors in $\Cmsc_{1}$.  
  Then, the following are true:
\begin{enumerate}
\item The dimension of $\Nmsc(\bar{\Ubf}_{\Ascr})$ is one.
\item For a nonzero vector $\vbf\in\Nmsc(\bar{\Ubf}_{\Ascr})$, the attack that modifies the sensor data from $\Cmsc_{1}$ by adding the corresponding entries in $\Ubf_{\Ascr}\vbf$ to the real data is equivalent to using $\alpha\cdot\abf^{*}$ as an attack vector, where $\alpha$ is a nonzero real number, and $\abf^{*}$ is an optimal solution to (\ref{eq:concept}) with $(\SmscA, \SmscF) = (\Cmsc_{1}, \Cmsc_{2})$. 
\end{enumerate}
\end{theorem}
\lisp

\begin{proof}
See Appendix~\ref{app:subspace_partial_framing}.
\end{proof}
\vspace{5pt}

Theorem~\ref{thm:subspace_partial_framing} implies that knowledge of a basis matrix of $\Rmsc(\Hbf_{\Ascr})$---the subspace of measurements from $\Smsco\setminus\Cmsc_{2}$---is sufficient for launching a data framing attack with $(\SmscA, \SmscF) = (\Cmsc_{1}, \Cmsc_{2})$.   Note that Theorem~\ref{thm:subspace_partial_framing} requires the same conditions as Theorem~\ref{thm:unobservable_partial}.   Therefore, for an attack on a power grid, the graph conditions in Corollary~\ref{cor:power_grid} can replace the conditions of Theorem~\ref{thm:subspace_partial_framing}.  

\subsection{Subspace data framing attack algorithm}\label{subsec:framing_data}

Theorem~\ref{thm:subspace_framing} and Theorem~\ref{thm:subspace_partial_framing} guarantee the sufficiency of subspace information in constructing data framing attacks.  
Similar to the data-driven algorithms for unobservable attacks, we can incorporate a subspace estimator and SVD to build a data-driven algorithm for data framing attacks.  

The data-driven framing attack with \emph{full} sensor observations receives sensor observations $\{\zbf_{1},\ldots,\zbf_{K}\}$ at $K$ different time instances and $(\SmscA,\SmscF)$ as inputs, and it has two small positive parameters $\epsilon_{1}$ and $\epsilon_{2}$ for thresholding rules.  Based on the QCQP formulation (\ref{eq:qcqp_2}), it  works as follows:
\begin{enumerate}
\item \textbf{Subspace estimation}: Based on $\{\zbf_{1},\ldots,\,\zbf_{K}\}$, calculate an estimate $\hat{\Ubf}\in\mbbR^{m\times n}$ of a basis matrix of $\Rmsc(\Hbf)$.
\item \textbf{Null space estimation}: Obtain $\hat{\Ubf}_{1}$ by removing the rows of $\hat{\Ubf}$ that correspond to the sensors in $\SmscA\cup\SmscF$.  Find an SVD of $\hat{\Ubf}_{1}$: $\hat{\Ubf}_{1} = \tilde{\Ubf}\tilde{\bm{\Lambda}}\tilde{\Vbf}^{T}$.  Let $\hat{\Vbf}$ denote the matrix consisting of the columns of $\tilde{\Vbf}$ whose corresponding singular values are less than $\epsilon_{1}$.  Let $\hat{\Ubf}_{\Ascr}\in\mbbR^{m\times n}$ be the matrix obtained from $\hat{\Ubf}$ by replacing the rows corresponding to the sensors \emph{not} in $\SmscA$ with zero row vectors.  Then, $\hat{\Ubf}_{\Ascr}\hat{\Vbf}$ is an estimate of a basis matrix of $\Rmsc(\Ubf_{1})\cap\Amsc$ in (\ref{eq:qcqp_2})\footnote{A basis matrix of $\Rmsc(\Ubf_{1})\cap\Amsc$ in (\ref{eq:qcqp_2}) can be found by noting that $\abf\in\Rmsc(\Ubf_{1})\cap\Amsc$ if and only if $\abf = \Ubf_{1}\ybf$ for some $\ybf\in\Nmsc(\Ubf_{2})$ where $\Ubf_{2}\in\mbbR^{(m-|\Smsc_{\text{\tiny A}}\cup\Smsc_{\text{\tiny F}}|) \times n}$ is a submatrix of $\Ubf$ obtained by removing the rows corresponding to the sensors in $\SmscA\cup\SmscF$.  In other words, given a basis matrix $\Bbf$ of $\Nmsc(\Ubf_{2})$, $\Ubf_{1}\Bbf$ is a basis matrix of $\Rmsc(\Ubf_{1})\cap\Amsc$.}. 
\item \textbf{QCQP parameter estimation}: Calculate 
\begin{equation}
\hat{\Wbf} \triangleq \Ibf - \hat{\Ubf}(\hat{\Ubf}^{T}\hat{\Ubf})^{-1}\hat{\Ubf}^{T}
\end{equation}
and $\hat{\bm{\Omega}}\in\mbbR^{m\times m}$, which is a diagonal matrix with 
\begin{equation}
\hat{\bm{\Omega}}_{ii} = \left\{\begin{array}{ll}
\sqrt{1/\hat{\Wbf}_{ii}} &\text{if $\hat{\Wbf}_{ii} > \epsilon_{2}$;}\\
0 & \text{if $\hat{\Wbf}_{ii} < \epsilon_{2}$.}
\end{array}
\right.
\end{equation} 
\item \textbf{QCQP}: 
Solve maximizing $\|\Ibf_{\Smsc_{\text{\tiny F}}}\hat{\bm{\Omega}} \hat{\Wbf}\hat{\Ubf}_{\Ascr}\hat{\Vbf}\ybf\|_{2}^{2}$
subject to $\|\hat{\Ubf}_{\Ascr}\hat{\Vbf}\ybf\|_{2}^{2} = 1$ and $\ybf\in\mbbR^{k}$, where $k$ is the number of columns of $\hat{\Vbf}$.  Let $\ybf^{*}$ denote the solution.

\item \textbf{Attack}:  Modify the sensor data from $\SmscA$ by adding the corresponding entries of $\eta\cdot\hat{\Ubf}_{\Ascr}\hat{\Vbf}\ybf^{*}$ to them, where $\eta\in\mbbR$ is a scaling factor to adjust the degree of perturbation.
\end{enumerate}

Based on Theorem~\ref{thm:subspace_partial_framing}, the data-driven framing attack with \emph{partial} sensor observations receives $(\Xmsco,\Smsco, \Cmsc_{1}, \Cmsc_{2})$ and $\{\tilde{\zbf}_{1},\ldots,\,\tilde{\zbf}_{K}\}$---the set of measurements from the sensors in $\Smsco\setminus\Cmsc_{2}$ at $K$ different time instances---as inputs and executes the following steps:
\begin{enumerate}
\item \textbf{Subspace estimation}: Based on $\{\tilde{\zbf}_{1},\ldots,\,\tilde{\zbf}_{K}\}$, calculate an estimate $\hat{\Ubf}_{\Ascr}\in\mbbR^{|\Smsc_{\text{\tiny O}}\setminus\Cmsc_{2}|\times |\Xmsc_{\text{\tiny O}}|}$ of a basis matrix of $\Rmsc(\Hbf_{\Ascr})$.

\item \textbf{Null space estimation}: Obtain $\hat{\Ubf}_{\cscr}$ by removing the rows of $\hat{\Ubf}_{\Ascr}$ that correspond to the sensors in $\Cmsc_{1}$.  Find an SVD of $\hat{\Ubf}_{\cscr}$: $\hat{\Ubf}_{\cscr} = \tilde{\Ubf}\tilde{\bm{\Lambda}}\tilde{\Vbf}^{T}$.  Let $\vbf$ denote the column of $\tilde{\Vbf}$ that corresponds to the smallest singular value ($\vbf$ is an estimate of a nonzero element of $\Nmsc(\bar{\Ubf}_{\Ascr})$ in Theorem~\ref{thm:subspace_partial_framing}.) 

\item \textbf{Attack}: Modify the sensor data from $\Cmsc_{1}$ by adding the corresponding entries of $\eta\cdot\hat{\Ubf}_{\Ascr}\vbf$ to them, where $\eta\in\mbbR$ is a scaling factor to adjust the degree of perturbation.
\end{enumerate}

\section{Numerical results}\label{sec:numerical_result}
In this section, simulations with benchmark power grids, the IEEE 14-bus network and the IEEE 118-bus network, demonstrate the performance of data-driven attacks.   The nonlinear measurement model (\ref{eq:AC_baddata}) and the nonlinear state estimator were employed to emulate practical power system state estimation.    The power system measurement model is briefly described in Appenidx~\ref{app:example}.  
As an attack performance metric, we used the $l_{2}$ norm of the mean state estimation error, \emph{i.e.}, $\mbbE[\|\hat{\xbf} - \xbf\|_{2}]$,
where $\hat{\xbf}$ is the state estimate, and $\xbf$ is the true state.

\subsection{Simulation methods}

In each Monte Carlo run, we used the nonlinear model (\ref{eq:AC_baddata}) to generate measurement vectors.
State vectors at different time points were assumed to be independent and identically distributed Gaussian random vectors with the mean equal to the operating states given in the IEEE 14-bus and 118-bus data\cite{IEEEParameter}.  The means are far from the nominal state that is generally used in a power system to obtain the linearized model (\ref{eq:DC_baddata}).   
The threshold of the bad data detector (\ie, the $J(\hat{\xbf})$-test) was set to satisfy the false alarm constraint 0.04.  

In each simulation scenario, we compared performance of three attack methods: an attack with full knowledge of $\Hbf$, a data-driven attack with full sensor observations, and a data-driven attack with partial sensor observations. 
For data-driven attacks, 1000 observations were used to estimate a basis matrix of the subspace of (either full or partial) measurements; the attacks employed the subspace estimator that uses the sample covariance matrix as described in Section~\ref{subsec:unobservable_data}.  
Both the 14-bus network and the 118-bus network were assumed to be fully measured; \ie, all bus injections and all line flows (in both directions for each line) were measured by sensors.
 
\begin{figure}[t!]
\centering
\includegraphics[width=.45\textwidth]{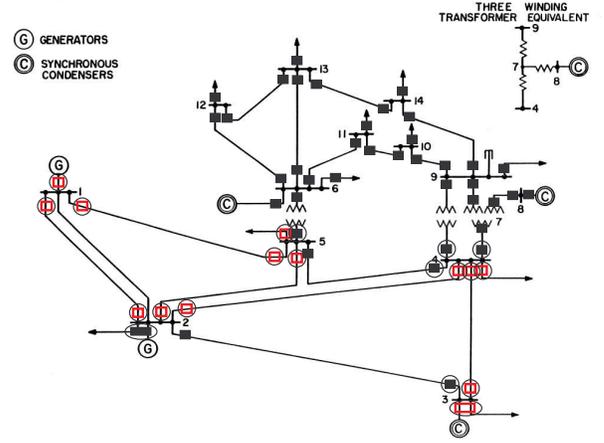}
\caption{IEEE 14-bus network:  The circled \emph{empty} rectangles represent the adversary sensors (\ie, the sensors in $\SmscA$).  The adversary with partial sensor observations can observe all the circled sensors.}
\label{fig:14bus}
\end{figure}

\subsection{Data-driven unobservable attack}
\subsubsection{IEEE 14-bus test}
In the IEEE 14-bus network, we considered an adversary controlling data from $(\bar{1})$, $(\bar{3})$, $(\bar{4})$, $(\bar{5})$, $(1, 2)$, $(2, 1)$, $(1, 5)$, $(5, 1)$, $(2, 5)$, $(5, 2)$, $(2, 4)$, $(4, 2)$, $(4, 3)$, and $(3, 4)$, as illustrated in Fig.~\ref{fig:14bus}: $(\bar{i})$ denotes the injection sensor at bus $i$, and $(i,j)$ denotes the line flow sensor measuring the power flow from $i$ to $j$.  Theorem~\ref{thm:feasibility_unobservable} and the spanning tree observability criterion \cite{Krumpholz&Clements&Davis:80PAS} imply that the adversary is capable of launching an unobservable attack (see Appendix~\ref{app:example}.)  In addition, the adversary sensor set is also a critical set,  and thus all possible unobservable attack vectors are aligned along the same direction (\ie, the dimension of $\Amsc\cap\Rmsc(\Hbf)$ is one.) 
 
An adversary with partial sensor observations was assumed to observe data from $(\bar{1})$, $(\bar{2})$, $(\bar{3})$, $(\bar{4})$, $(\bar{5})$, $(1, 2)$, $(2, 1)$, $(1, 5)$, $(5, 1)$, $(2, 5)$, $(5, 2)$, $(2, 4)$, $(4, 2)$, $(3, 4)$, $(4, 3)$, $(4, 5)$, $(3, 2)$, $(5, 6)$, $(4, 7)$, and $(4, 9)$.  
In this setting, the spanning tree observability criterion can be used to verify that the conditions of Theorem~\ref{thm:unobservable_partial} are satisfied (see Appendix~\ref{app:example},) and thus an adversary with partial observations can construct an unobservable attack under the linearized model assumption.  

\begin{figure}[t!]
\centering
\includegraphics[width=.45\textwidth]{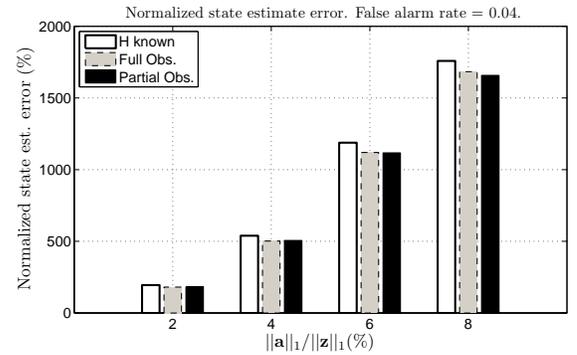}
\caption{Unobservable attacks on the 14-bus network: the sensor SNR is 46dB.  Attacks with the relative attack magnitudes 2, 4, 6, and 8 $\%$ were tested.  For each scenario, 1,000 Monte Carlo runs are used.} 
\label{fig:14bus_unobservable}
\end{figure}

Fig.~\ref{fig:14bus_unobservable} shows the performance of unobservable attacks, especially the plot of the \emph{normalized} state estimation error versus the relative attack magnitude ($\|\abf\|_{1}/\|\zbf\|_{1}$).  
The mean state estimation errors are normalized with respect to the mean estimation error under the non-attack scenario.  
Both data-driven attacks performed as well as the attack with knowledge of $\Hbf$.  The results indicate that even in a practical {nonlinear} power system, the data-driven attacks designed based on the linear model can perform well, and partial sensor observations can provide sufficient information for designing an unobservable attack.

\subsubsection{IEEE 118-bus test}
\begin{figure}[t!]
\centering
\includegraphics[width=.46\textwidth]{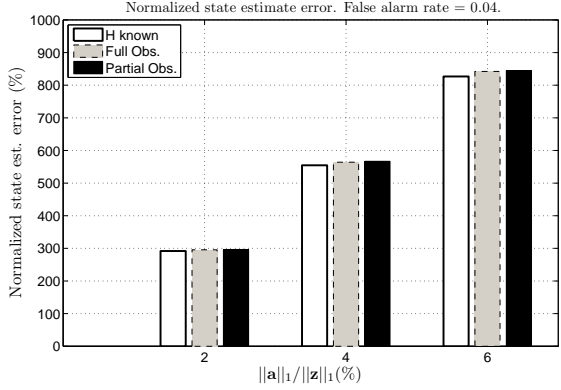}
\caption{Unobservable attacks on the 118-bus network: the sensor SNR is 46dB.  Attacks with the relative attack magnitudes 2, 4, and 6 $\%$ were tested.  For each scenario, 200 Monte Carlo runs are used.} 
\label{fig:118bus_unobservable}
\end{figure}

In the IEEE 118-bus simulation, we considered unobservable attacks discussed in the example in Fig.~\ref{fig:118bus_part} of Section~\ref{subsec:unobservable_partial}.  Fig.~\ref{fig:118bus_unobservable} shows the plots of the normalized state estimation error versus the relative attack magnitude.  Three methods resulted in almost the same degree of perturbation on the state estimate.  The results demonstrate that observing data from a \emph{small fraction} of sensors can be sufficient for launching an unobservable attack on a large system; only about 2 percent of sensors need to be observed.  

\subsection{Data-driven framing attack}
\subsubsection{IEEE 14-bus test}
For data framing attacks, we considered an adversary who controls $(\bar{4})$, $(1, 5)$, $(5, 1)$, $(5, 2)$, $(4, 2)$, $(4, 3)$, and $(3, 4)$, and frames $(\bar{1})$, $(\bar{3})$, $(\bar{5})$, $(1, 2)$, $(2, 1)$, $(2, 5)$, and $(2, 4)$ as sources of bad data.  Under this setting, an adversary cannot launch an unobservable attack.     
An adversary with partial observations was assumed to observe data from $(\bar{2})$, $(\bar{4})$, $(1, 5)$, $(5, 1)$, $(5, 2)$, $(4, 2)$, $(3, 4)$, $(4, 3)$, $(4, 5)$, $(3, 2)$, $(5, 6)$, $(4, 7)$, and $(4, 9)$.  
This setting satisfies the conditions of Theorem~\ref{thm:subspace_partial_framing} and enables an adversary with partial sensor observations to launch a data framing attack under the linearized model assumption (see Appendix~\ref{app:example}.)
 
Fig.~\ref{fig:14bus_framing} shows the plots of the normalized state estimation error versus the relative attack magnitude.  
The results show that even when an unobservable attack is not feasible, an adversary may exploit the idea of data framing to perturb the state estimate by an arbitrary degree. 
Furthermore, the results indicate that partial sensor observations are sufficient for designing a data framing attack.

\begin{figure}[t!]
\centering
\includegraphics[width=.45\textwidth]{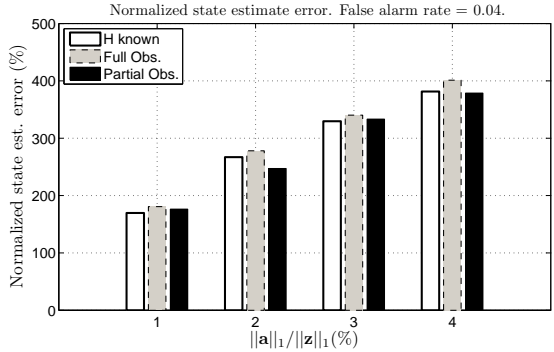}
\caption{Data framing attacks on the 14-bus network: the sensor SNR is 46dB.  Attacks with the relative attack magnitudes 1, 2, 3, and 4 $\%$ were tested.  For each scenario, 1,000 Monte Carlo runs are used.} 
\label{fig:14bus_framing}
\end{figure}

\subsubsection{IEEE 118-bus test}
We considered an adversary attacking the part of the 118-bus network illustrated in Fig.~\ref{fig:118bus_part}.   The adversary was assumed to control $(114, 115)$, $(115, 114)$, and $(27, 115)$, and frame $(\bar{114})$, $(\bar{115})$, $(\bar{27})$, and $(115, 27)$ as sources of bad data.   
 An adversary with partial sensor observations was assumed to observe data from the circled sensors in Fig.~\ref{fig:118bus_part} except $(\bar{114})$, $(\bar{115})$, $(\bar{27})$, and $(115, 27)$.  
The graph conditions of Corollary~\ref{cor:power_grid} are satisfied, and thus  
an adversary with partial observations is capable of launching a data framing attack under the linearized model assumption.

\begin{figure}[t!]
\centering
\includegraphics[width=.46\textwidth]{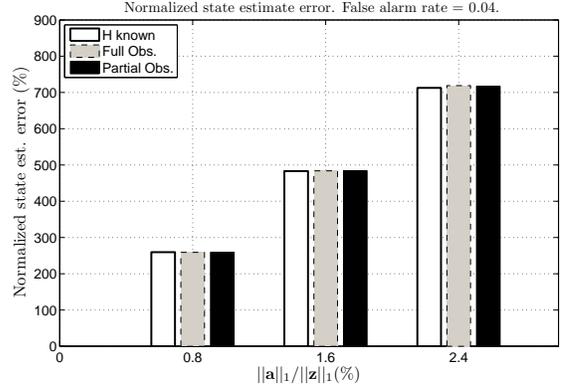}
\caption{Data framing attacks on the 118-bus network: the sensor SNR is 46dB.  Attacks with the relative attack magnitudes 0.8, 1.6, and 2.4 $\%$ were tested.  For each scenario, 200 Monte Carlo runs are used.}  
\label{fig:118bus_framing}
\end{figure}

Fig.~\ref{fig:118bus_framing} shows the plots of the normalized state estimation error versus the relative attack magnitude.  The results demonstrate the sufficiency of partial sensor observations for designing a data framing attack in a large network.   

\section{Conclusions}\label{sec:conclusion_frame}
This paper presents subspace methods of data attacks on state estimators of cyber physical systems.   
By exploiting the fact that subspace information of measurements is sufficient for designing attacks, we devised data-driven attacks that can be launched based on partial sensor observations. 
The numerical results demonstrated that the data-driven attacks are as efficient as the attacks based on full system information.

Our results demonstrate that one should not presumably underestimate the ability of an adversary even when system information is secure from the adversary.  
Even a leak of a small fraction of certain sensor measurements may provide enough data, upon which state attacks can be constructed.    

Most countermeasures in the literature focused on protecting certain sensor data from adversarial modification via data authentication, while assuming that system parameters are known to adversaries (\eg, \cite{Bobba&etal:10SCS, Kim&Poor:2011TSG, Giani2013:TSG, KimTong:2013SGC}).  
In case that system parameter information is kept secure, our results demonstrate that not only the ability to modify data but also the ability to observe data are critical to an adversary.  Therefore, as a countermeasure, on top of a data authentication strategy, one can strategically enhance data encryption and access control protocols to limit the set of data an adversary may eavesdrop.

\begin{appendices}

\section{Proof of Theorem~\ref{thm:feasibility_unobservable}}\label{app:Kosut}
Let $\bar{\Hbf}$ denote the measurement matrix after the sensors in $\SmscA$ are removed; \ie, $\bar{\Hbf}$ is obtained from $\Hbf$ by removing the rows corresponding to the adversary sensors.  
Then, $\Hbf\ybf$ is in $\Amsc$ if and only if $\ybf$ is in $\Nmsc(\bar{\Hbf})$---the null space of $\bar{\Hbf}$.
This implies that an unobservable attack is feasible if and only if $\bar{\Hbf}$ does not have full column rank (\ie, $\Nmsc(\bar{\Hbf})$ has a nonzero dimension.)  \endproof

\section{Proof of Theorem~\ref{thm:unobservable_subspace}}\label{app:unobservable_subspace}
The columns of $\bar{\Ubf}$ span $\Rmsc(\bar{\Hbf})$.
In addition, because $\bar{\Ubf}$ and $\bar{\Hbf}$ have the same number of columns, $\bar{\Ubf}$ does not have full column rank if and only if $\bar{\Hbf}$ does not have full column rank.  Therefore, Theorem~\ref{thm:feasibility_unobservable} implies that an unobservable attack is feasible if and only if $\bar{\Ubf}$ does not have full column rank.


Suppose that an unobservable attack is feasible.  Then, $\bar{\Ubf}$ is rank deficient, and we can find a nonzero vector $\vbf\in\Nmsc(\bar{\Ubf})$.   With $\abf \triangleq \Ubf\vbf$,
$\abf$ is in $\Amsc$ because $\Ubf\vbf$ has zero entries for the sensors \emph{not} in $\SmscA$ (\ie, $\bar{\Ubf}\vbf = \textbf{0}$).  
In addition, there exists an invertible matrix $\Bbf\in\mbbR^{n\times n}$ such that $\Hbf = \Ubf\Bbf$, and $\Ubf = \Hbf\Bbf^{-1}$, because $\Hbf$ has full column rank.  Therefore, $\Ubf\vbf = \Hbf(\Bbf^{-1}\vbf)$, and thus $\abf$ is an unobservable attack vector.  \endproof

\section{Proof of Theorem~\ref{thm:unobservable_partial}}\label{app:unobservable_partial}
Let $\bar{\Hbf}$ denote the submatrix of $\Hbf$ obtained by removing the rows corresponding to the sensors in $\Cmsc$.  Then, $\Nmsc(\bar{\Hbf})$ is not null due to the third assumption.  Let $\ybf$ denote a nonzero vector in $\Nmsc(\bar{\Hbf})$ and $\ybf_{\oscr}$ denote a subvector of $\ybf$ obtained by retaining only the rows corresponding to the state variables in $\Xmsco$.  In addition, let $\Hbf_{\text{s}}$ denote a submatrix of $\Hbf_{\oscr}$ obtained by retaining only the columns corresponding to the state variables in $\Xmsco$ (note that all the other columns of $\Hbf_{\oscr}$ are zero vectors.)  And, $\bar{\Hbf}_{\text{s}}$ denotes a submatrix of $\Hbf_{\text{s}}$ obtained by removing the rows corresponding to the sensors in $\Cmsc$.  

First, note that 
$\abf_{\oscr}\in\Amsc_{\oscr}$ if and only if $\abf_{\oscr} = \Hbf_{\text{s}}\pbf$ for some $\pbf\in\Nmsc(\bar{\Hbf}_{\text{s}})$.  In addition, because $\Cmsc$ is a critical set with respect to $(\Smsco,\Xmsco)$, $\Nmsc(\bar{\Hbf}_{\text{s}})$ has dimension one.
Note that $\bar{\Hbf}_{\text{s}}\ybf_{\oscr} = \textbf{0}$ whereas ${\Hbf}_{\text{s}}\ybf_{\oscr} \neq \textbf{0}$.  This implies that $\ybf_{\oscr}\neq \textbf{0}$, and 
$\{\ybf_{\oscr}\}$ is a basis of $\Nmsc(\bar{\Hbf}_{\text{s}})$. 
Therefore, $\{\Hbf_{\text{s}}\ybf_{\oscr}\}$ is a basis of $\Amsc_{\oscr}$.

Therefore, for any nonzero $\abf_{\oscr}\in\Amsc_{\oscr}$, there exists a nonzero $\alpha\in\mbbR$ such that
$\abf_{\oscr} = \alpha\cdot\Hbf_{\text{s}}\ybf_{\oscr}$.  Furthermore, $\Hbf_{\text{s}}\ybf_{\oscr} = \Hbf_{\oscr}\ybf$ implies that 
\begin{equation}
\abf_{\oscr} = \alpha\cdot\Hbf_{\oscr}\ybf.
\end{equation}
In addition, $\bar{\Hbf}\ybf = \textbf{0}$ implies that the attack that modifies the data from $\Cmsc$ by adding the corresponding entries of $\abf_{\oscr}$ to the actual data is equivalent to using $\alpha\cdot\Hbf\ybf$ as an attack vector, which is unobservable.  So, the attack is unobservable.
\endproof

\section{Proof of Theorem~\ref{thm:subspace_framing}}\label{app:subspace_framing}
The normalized residues in the first iteration are affected by the attack $\abf$ as follows:
\begin{equation}
\tilde{\rbf} = \bm{\Omega}\Wbf(\zbf + \abf) = \bm{\Omega}\Wbf\ebf + \bm{\Omega}\Wbf\abf,
\end{equation}
which can be derived from (\ref{eq:normalize}) and (\ref{eq:r_DC}).  Note that $(\bm{\Omega}\Wbf\ebf)_{i}$ follows a standard normal distribution (due to the normalization) if $\{i\}$ is not a critical set; $(\bm{\Omega}\Wbf\ebf)_{i}$ is zero otherwise.
Therefore, $\tilde{\rbf}_{i}$ follows the normal distribution $\Nc((\bm{\Omega}\Wbf\abf)_{i}, 1)$ if $\{i\}$ is not a critical set; otherwise, $\tilde{\rbf}_{i}$ is equal to $(\bm{\Omega}\Wbf\abf)_{i}$.

Therefore, the expected energy of the normalized residues at $\SmscF$ in the presence of the attack $\abf$ is
\begin{equation}
\mbbE\left[\sum_{i\in\Smsc_{\text{\tiny F}}}(\tilde{r}_{i})^{2}\right] = \sum_{i\in\Smsc_{\text{\tiny F}}}(\bm{\Omega} \Wbf\abf)_{i}^{2} + C = \|\Ibf_{\Smsc_{\text{\tiny F}}}\bm{\Omega} \Wbf\abf\|_{2}^{2} + C,
\end{equation}
where $C$ is the number of sensors in $\SmscF$ that do not form a single element critical set.

Consequently, a solution to (\ref{eq:concept}) is also a solution to the following problem, and vice versa:
\begin{equation}\label{eq:qcqp}
\begin{array}{ll}
\max_{\abf} & \|\Ibf_{\Smsc_{\text{\tiny F}}}\bm{\Omega} \Wbf\abf\|_{2}^{2}\\
\text{subj.} & \|\abf\|_{2}^{2} = 1, ~~ \abf\in\Rmsc(\Hbf_{1})\cap\Amsc,
\end{array}
\end{equation}

The theorem statements follow from the following observations: $\Wbf$ is equal to $\tilde{\Wbf}$ as both are orthogonal projections on the same space, and $\Rmsc(\Hbf_{1})$ is equivalent to $\Rmsc(\Ubf_{1})$.
\endproof

\section{Proof of Theorem~\ref{thm:subspace_partial_framing}}\label{app:subspace_partial_framing}
Let $\bar{\Hbf}$ denote the submatrix of $\Hbf$ obtained by removing the rows corresponding to the sensors in $\Cmsc$.  
First, from the proof procedure of Theorem~\ref{thm:unobservable_partial}, one can derive that the dimension of $\Nmsc(\bar{\Hbf})$ is one. This implies that $\Cmsc$ contains exactly one critical set.  Because, if there were more than one critical sets included in $\Cmsc$, $\Nmsc(\bar{\Hbf})$ should have a dimension larger than one.

Because $\SmscA\cup\SmscF = \Cmsc$ contains exactly one critical set, the dimension of $\Rmsc(\Hbf_{1})\cap\Amsc$ in (\ref{eq:concept}) is one.  This can be seen as follows.   
The dimension of $\Rmsc(\Hbf_{1})\cap\Amsc$ in (\ref{eq:concept}) is equal to the dimension of $\Nmsc(\Hbf_{2})$ where $\Hbf_{2}$ is the matrix obtained from $\Hbf$ by removing the rows corresponding to the sensors in $\SmscA\cup\SmscF$.  And, the fact that $\SmscA\cup\SmscF$ contains exactly one critical set implies that the rank of $\Hbf_{2}$ is $n-1$, and thus the dimension of $\Nmsc(\Hbf_{2})$ is $1$.  

Therefore, (\ref{eq:concept}) has only two feasible points, and they give the same objective function values.  In particular, a solution to (\ref{eq:concept}) is the direction given by $\Hbf_{1}\Delta\xbf$ where $\Delta\xbf$ is a nonzero vector in $\Nmsc(\Hbf_{2})$ (see \cite{KimTongThomas:2013ArXiv} for more detailed arguments.)  

The first and second conditions of Theorem~\ref{thm:unobservable_partial}, which are assumed to hold, imply that the dimension of $\Nmsc(\bar{\Ubf}_{\Ascr})$ is one. 
In addition, it can be seen from Corollary~\ref{cor:local_unobservable} that the second statement is true for $\abf^{*} = \Hbf_{1}\Delta\xbf$ and some nonzero $\alpha$.
\endproof

\section{Power grid measurement model and observability}\label{app:example}

In this section, we briefly describe the power system measurement model and the spanning-tree observability criterion in \cite{Krumpholz&Clements&Davis:80PAS}.  The spanning-tree observability criterion results in Corollary~\ref{cor:power_grid} from Theorem~\ref{thm:unobservable_partial}.  For more details about power system models, see \cite{Abur&Exposito:book}.

The power system state is defined as the vector of voltage magnitudes and phase angles at all buses except a reference bus, which is an arbitrary bus whose voltage phase angle is set to zero:
\begin{equation}
\xbf = [V_{1}~V_{2}~\cdots~V_{n}~\theta_{2}~\cdots~\theta_{n}]^{T}
\end{equation}
where $V_{i}$ and $\theta_{i}$ denote the voltage magnitude and phase angle at bus $i$ respectively, and bus 1 is set as the reference bus. 

 We consider two types of legacy sensors: line flow sensors and bus injection sensors\footnote{Other types of sensors (\eg, phasor measurement units) can also be considered.  We impose this restriction merely to facilitate clearer presentation.}.  
The line flow from bus $i$ to bus $j$ is a complex quantity related to the system state by 
\begin{equation}
P_{ij} + \textrm{j}\cdot Q_{ij} = V_{i}e^{\textrm{j}\theta_{i}}\cdot\left(\dfrac{V_{i}e^{\textrm{j}\theta_{i}} -V_{j}e^{\textrm{j}\theta_{j}}}{Z_{ij}}\right)^{*}
\end{equation}
where $P_{ij}\in\mbbR$ and $Q_{ij}\in\mbbR$ are real and imaginary parts of the line flow respectively, $Z_{ij}$ is the impedance of the line $\{i,j\}$, and $X^{*}$ denotes the complex conjugate of $X$.  The bus injection at bus $i$ is the sum of all outgoing line flows from bus $i$.

For computational benefits, the above nonlinear relation is often linearized at the nominal operating point where all bus voltage magnitudes are equal to $1$ p.u., and all bus voltage phase angles are equal to zero.   This linearization decouples the relation such that the real part of measurements depends only on the voltage phase angles, and the imaginary part depends only on the voltage magnitudes.

The linearized relation between the real part of measurements and the voltage phase angles---the so-called DC model---is often used to analyze power system observability.  
In the DC model (\ref{eq:DC_baddata}), the state $\xbf$ is defined as the vector of voltage {phase angles} at all buses except the reference bus:
\begin{equation}
\xbf = [\theta_{2}~\theta_{3}~\cdots~\theta_{n}]^{T}.
\end{equation} 
The measurement matrix $\Hbf$ depends on the topology and line impedance\footnote{To describe the entries of $\Hbf$, we consider a noiseless measurement vector $\zbf = \Hbf\xbf$ for simplicity. Suppose that the $k$th entry of $\zbf$ is a measurement from a line flow sensor measuring the line flow from bus $i$ to $j$.  Then, if the line is \emph{connected}, $z_{k} = B_{ij}(\theta_{i} - \theta_{j})$, where $B_{ij}$ is the susceptance of the line; if the line is \emph{not} connected, $z_{k} = 0$.  In case that $z_k$ corresponds to an injection sensor at bus $i$, $z_{k}$ is the sum of all the outgoing line flows from bus $i$.}.  

The power system is observable if and only if $\Hbf$ has full column rank\cite{Krumpholz&Clements&Davis:80PAS}.  Verifying this rank condition seems to require knowledge of the line impedance.  However, Krumpholz \emph{et al.} \cite{Krumpholz&Clements&Davis:80PAS} showed that system observability can be determined purely based on the topology and sensor locations.  In particular, Krumpholz \emph{et al.} \cite{Krumpholz&Clements&Davis:80PAS} showed that a system is observable 
 if and only if there exists a way to assign each injection sensor to any of the lines that are incident to the bus where the sensor is located such that there exists a spanning tree of the topology having at least one sensor (an assigned injection or line flow sensor) on each edge of the tree (see Corollary 2 in \cite{Krumpholz&Clements&Davis:80PAS}.)

The spanning tree criterion can also be used to check whether the state variables in $\Xmsco$ are observable with respect to $\Smsco$ (we use the notations in Section~\ref{subsec:unobservable_partial}.)  Without loss of generality, we assume that $\Smsco$ contains an injection sensor on the reference bus or a line flow sensor on a line incident to the reference bus\footnote{Note that we can choose the reference bus such that this condition holds.}.  Then, we can simply apply the spanning tree criterion to the reduced network for $\Smsco$ (see Section~\ref{subsec:unobservable_partial} for the definition of a reduced network.)  
The state variables in $\Xmsco$ are observable with respect to $\Smsco$ if and only if it is possible to assign injection sensors in $\Smsco$ to their neighboring lines such that  a spanning tree of the reduced network with at least one sensor in $\Smsco$ on every edge exists.  

\end{appendices}



\begin{thebibliography}{10}
\providecommand{\url}[1]{#1}
\csname url@samestyle\endcsname
\providecommand{\newblock}{\relax}
\providecommand{\bibinfo}[2]{#2}
\providecommand{\BIBentrySTDinterwordspacing}{\spaceskip=0pt\relax}
\providecommand{\BIBentryALTinterwordstretchfactor}{4}
\providecommand{\BIBentryALTinterwordspacing}{\spaceskip=\fontdimen2\font plus
\BIBentryALTinterwordstretchfactor\fontdimen3\font minus
  \fontdimen4\font\relax}
\providecommand{\BIBforeignlanguage}[2]{{%
\expandafter\ifx\csname l@#1\endcsname\relax
\typeout{** WARNING: IEEEtran.bst: No hyphenation pattern has been}%
\typeout{** loaded for the language `#1'. Using the pattern for}%
\typeout{** the default language instead.}%
\else
\language=\csname l@#1\endcsname
\fi
#2}}
\providecommand{\BIBdecl}{\relax}
\BIBdecl

\bibitem{Lee2008UCB_TR}
\BIBentryALTinterwordspacing
E.~A. Lee, ``Cyber physical systems: Design challenges,'' EECS Department,
  University of California, Berkeley, Tech. Rep. UCB/EECS-2008-8, Jan 2008.
  [Online]. Available:
  \url{http://www.eecs.berkeley.edu/Pubs/TechRpts/2008/EECS-2008-8.html}
\BIBentrySTDinterwordspacing

\bibitem{Huang2012:SPMag}
Y.-F. Huang, S.~Werner, J.~Huang, N.~Kashyap, and V.~Gupta, ``State estimation
  in electric power grids: Meeting new challenges presented by the requirements
  of the future grid,'' \emph{IEEE Signal Processing Magazine}, vol.~29, no.~5,
  pp. 33--43, Sept 2012.

\bibitem{Cardenas2009DHSWorkshop}
A.~Cardenas, S.~Amin, B.~Sinopoli, A.~Giani, A.~Perrig, and S.~S. Sastry,
  ``Challenges for securing cyber physical systems,'' in \emph{Workshop on
  Future Directions in Cyber-physical Systems Security}.\hskip 1em plus 0.5em
  minus 0.4em\relax DHS, July 2009.

\bibitem{Hull:2012PEMag}
J.~Hull, H.~Khurana, T.~Markham, and K.~Staggs, ``Staying in control:
  Cybersecurity and the modern electric grid,'' \emph{IEEE Power and Energy
  Magazine}, vol.~10, no.~1, pp. 41--48, 2012.

\bibitem{INL2011}
``{Vulnerability Analysis of Energy Delivery Control Systems},'' {Idaho
  National Laboratory}, September 2011, {INL/EXT-10-18381}.

\bibitem{Liu:2009CCS}
Y.~Liu, P.~Ning, and M.~K. Reiter, ``False data injection attacks against state
  estimation in electric power grids,'' in \emph{Proceedings of the 16th ACM
  conference on Computer and communications security}, 2009, pp. 21--32.

\bibitem{Bobba&etal:10SCS}
R.~B. Bobba, K.~M. Rogers, Q.~Wang, H.~Khurana, K.~Nahrstedt, and T.~J.
  Overbye, ``{Detecting false data injection attacks on dc state estimation},''
  in \emph{First Workshop on Secure Control Systems,CPSWEEK 2010}, Stockholm,
  Sweeden, Apr 2010.

\bibitem{Kosut11}
O.~Kosut, L.~Jia, R.~J. Thomas, and L.~Tong, ``Malicious data attacks on the
  smart grid,'' \emph{IEEE Transactions on Smart Grid}, vol.~2, no.~4, pp. 645
  --658, Dec. 2011.

\bibitem{Kim&Poor:2011TSG}
T.~Kim and H.~Poor, ``Strategic protection against data injection attacks on
  power grids,'' \emph{IEEE Transactions on Smart Grid}, vol.~2, no.~2, pp. 326
  --333, june 2011.

\bibitem{Bi&Zhang:2011Globecom}
S.~Bi and Y.~Zhang, ``Defending mechanisms against false-data injection attacks
  in the power system state estimation,'' in \emph{2011 IEEE GLOBECOM
  Workshops}, Houston, TX, USA., Dec 2011.

\bibitem{GianiEtal:2011SGC}
A.~Giani, E.~Bitar, M.~Garcia, M.~McQueen, P.~Khargonekar, and K.~Poolla,
  ``Smart grid data integrity attacks: characterizations and countermeasures,''
  in \emph{2011 IEEE International Conference on Smart Grid Communications
  (SmartGridComm)}, Oct 2011, pp. 232--237.

\bibitem{KimTong:2013SGC}
J.~Kim and L.~Tong, ``On phasor measurement unit placement against state and
  topology attacks,'' in \emph{IEEE International Conference on Smart Grid
  Communications}, Oct. 2013.

\bibitem{EsmalifalakEtal:2011SGC}
M.~Esmalifalak, H.~Nguyen, R.~Zheng, and Z.~Han, ``Stealth false data injection
  using independent component analysis in smart grid,'' in \emph{IEEE
  International Conference on Smart Grid Communications}, Oct. 2011, pp.
  244--248.

\bibitem{Rahman:2012Globecom}
M.~Rahman and H.~Mohsenian-Rad, ``False data injection attacks with incomplete
  information against smart power grids,'' in \emph{IEEE Global Communications
  Conference (GLOBECOM)}, Dec. 2012.

\bibitem{Stoica1989TASSP}
P.~Stoica and A.~Nehorai, ``Music, maximum likelihood, and cramer-rao bound,''
  \emph{IEEE Transactions on Acoustics, Speech and Signal Processing}, vol.~37,
  no.~5, pp. 720--741, 1989.

\bibitem{Pezeshki2008TSP}
A.~Pezeshki, B.~Van~Veen, L.~Scharf, H.~Cox, and M.~Nordenvaad, ``Eigenvalue
  beamforming using a multirank mvdr beamformer and subspace selection,''
  \emph{IEEE Transactions on Signal Processing}, vol.~56, no.~5, pp.
  1954--1967, 2008.

\bibitem{Viberg1995Automatica}
M.~Viberg, ``Subspace-based methods for the identification of linear
  time-invariant systems,'' \emph{Automatica}, vol.~31, no.~12, pp. 1835 --
  1851, 1995.

\bibitem{IEEEParameter}
\BIBentryALTinterwordspacing
``{Power Systems Test Case Archive}.'' [Online]. Available: \url{{\tt
  http://www.ee.washington.edu/research/pstca/}}
\BIBentrySTDinterwordspacing

\bibitem{Krumpholz&Clements&Davis:80PAS}
G.~R. Krumpholz, K.~A. Clements, and P.~W. Davis, ``{Power system
  observability: a practical algorithm using network topology},'' \emph{IEEE
  Trans. Power Apparatus and Systems}, vol.~99, no.~4, pp. 1534--1542, July
  1980.

\bibitem{Giani2013:TSG}
A.~Giani, E.~Bitar, M.~Garcia, M.~McQueen, P.~Khargonekar, and K.~Poolla,
  ``Smart grid data integrity attacks,'' \emph{IEEE Transactions on Smart
  Grid}, vol.~4, no.~3, pp. 1244--1253, Sept 2013.

\bibitem{KimTong:13JSAC}
J.~Kim and L.~Tong, ``{On topology attack of a smart grid: undetectable attacks
  and countermeasures},'' \emph{IEEE Journal on Selected Areas in
  Communications}, vol.~31, no.~7, July 2013.

\bibitem{Sandberg&Teixerira&Johansson:10SCS}
H.~Sandberg, A.~Teixeira, and K.~H. Johansson, ``{On security indices for state
  estimators in power networks},'' in \emph{First Workshop on Secure Control
  Systems,CPSWEEK 2010}, Stockholm, Sweeden, Apr 2010.

\bibitem{KimTongThomas:2013ArXiv}
J.~{Kim}, L.~{Tong}, and R.~J. {Thomas}, ``{Data Framing Attack on State
  Estimation},'' \emph{ArXiv e-prints, {\tt arXiv:1310.7616}}, Apr. 2014, to
  appear in \emph{IEEE Journal on Selected Areas in Communications}.

\bibitem{MoSinopoli2010:CPSWEEK}
Y.~Mo and B.~Sinopoli, ``False data injection attacks in control systems,'' in
  \emph{First Workshop on Secure Control Systems, CPS Week}, 2010.

\bibitem{Pasqualetti2013:TAC}
F.~Pasqualetti, F.~Dorfler, and F.~Bullo, ``Attack detection and identification
  in cyber-physical systems,'' \emph{IEEE Transactions on Automatic Control},
  vol.~58, no.~11, pp. 2715--2729, Nov 2013.

\bibitem{Abur&Exposito:book}
A.~Abur and A.~G. Exp\'{o}sito, \emph{Power System State Estimation: Theory and
  Implementation}.\hskip 1em plus 0.5em minus 0.4em\relax CRC, 2000.

\bibitem{Jia2014:TPS}
L.~Jia, J.~Kim, R.~Thomas, and L.~Tong, ``Impact of data quality on real-time
  locational marginal price,'' \emph{IEEE Transactions on Power Systems},
  vol.~29, no.~2, pp. 627--636, March 2014.

\bibitem{Handschin&Schweppe&Kohlas&Feichter:75TPAS}
E.~Handschin, F.~C. Schweppe, J.~Kohlas, and A.~Fiechter, ``{Bad data analysis
  for power system state estimation},'' \emph{IEEE Trans. Power Apparatus and
  Systems}, vol. PAS-94, no.~2, pp. 329--337, Mar/Apr 1975.

\bibitem{Srivastava2000TSP}
A.~Srivastava, ``{A Bayesian Approach to Geometric Subspace Estimation},''
  \emph{IEEE Transactions on Signal Processing}, vol.~48, no.~5, pp.
  1390--1400, May 2000.

\bibitem{Smith2005TSP}
S.~T. Smith, ``{Covariance, Subspace, and Intrinsic Cramer Rao Bounds},''
  \emph{IEEE Transactions on Signal Processing}, vol.~53, no.~5, pp.
  1610--1630, May 2005.

\end{thebibliography}
\end{document}